\documentclass[prd,tightenlines,twocolumn]{revtex4}
\usepackage{amsmath}
\usepackage{amssymb}
\usepackage{graphicx}
\usepackage{bm}
\usepackage{enumerate}
\usepackage{color}
\newcommand{\be}{\begin{equation}}
\newcommand{\ee}{\end{equation}}
\newcommand{\bea}{\begin{eqnarray}}
\newcommand{\eea}{\end{eqnarray}}
\newcommand{\ba}{\begin{eqnarray}}
\newcommand{\ea}{\end{eqnarray}}

\begin{document}

\title{Comments on the nucleon spin composition and diquark correlations}
\author{ 
 Edward Shuryak }

\affiliation{Department of Physics and Astronomy, \\ Stony Brook University,\\
Stony Brook, NY 11794, USA}


\begin{abstract} 
We discuss possible relation between
recent lattice results on the flavor composition of the nucleon spin 
and the topology-induced quark correlations.
 \end{abstract}

\maketitle
 The  so called ``nucleon spin problem" is a part of a wider issue, of understanding
of the hadronic structure functions. Multiple experiments have revealed 
that, contrary to perturbative expectations, the PDFs of the  ``sea" quarks and antiquarks 
of a polarized nucleon are rather strongly polarized, both in spin and even the flavor (the distributions
of $\bar u, \bar d,\bar s$ are $not$ the same). 

Lattice simulations of QCD with realistic quark masses and inclusion of disconnected diagrams, which only recently became feasible,
have provided 
 further interesting observations. These comments were, in particular, triggered 
 by Ref. \cite{Alexandrou:2017oeh}, which in particular compared the flavor contributions of $u,d,s$ quarks and gluons
 to two observables, the nucleon spin and its total momentum. While in the latter case
 the contributions of the $d$ quark to nucleon's total momentum is, as expected, close to half of that of $u$,
 a completely different flavor distribution is found for the spin:
 the contribution of the $d$  and $s$ quarks are much smaller, being $11\pm 8\%$ and $9\pm 4\%$ respectively.
 It is also interesting that the gluon contribution to both is, on the contrary, the same, $27\pm 3\%$.

The main point of these comments is to suggest an  explanation for  the $d$  and $s$ quarks failing to
contribute to the nucleon spin: namely that they mostly appear in the nucleon wave function in form 
of {\em spin zero diquarks} , $ud$ and $us$.

Existence of diquark correlations in baryons and their parameters has been discussed in  vast literature
since the beginning of $SU(3)_f$ symmetry in 1960's. During the last decade experimental discoveries of multiquark
states with heavy $c,b$ quarks had initiated large activity in hadronic spectroscopy. Many of those papers
also use diquarks as a tool, say reducing the number of bodies from five for pentaquark to three. These comments are
not aimed at covering any of that, and in particularly not discussing diquarks containing heavy quarks.

Our focus is on the dynamical origins of diquark correlations, their relation to chiral symmetry breaking,
and thus dependence on the quark (pion) mass.
The points to be made below are all known but
  scattered in multiple publications. Therefore, we remind
here the main points, once again. 

Starting with the kinematics, it is obvious from Fermi statistics that scalar (spin zero) diquarks
must be $antisymmetric$ in flavor, that is  $ud$ and $us$ but not $uu$, while the vector ones (spin one) must be flavor symmetric.  

Let me then present the symmetry argument, from
\cite{Rapp:1997zu}, based on continuation in the number of colors $N_c$. In the massless two-color $N_c=2$ QCD  there exist additional Pauli-Gursey symmetry, which mixes quarks with anti-quarks. As a result,  diquarks (which are of course color singlets) are degenerate with the corresponding mesons. Chiral symmetry breaking is in fact such that scalar diquarks are Goldstone bosons, like pions, and thus must also be $massless$,  small size etc. The
vector diquarks are  degenerate with the vector $\rho,\omega$ mesons. Now, proceeding from $N_c=2$ to  $N_c=3$
QCD, one expect to see continuity, in form of not massless but deeply bound scalar diquarks. 
So to say, the scalar diquark is no longer the pion twin, but remains its half-brother.
Strong quark pairing in the scalar channel was then recognized as a basis for {\em color superconductivity}
in dense quark matter. 

To proceed to more quantitative estimates one needs to have certain dynamical information about the
binding effects. We will not repeat here discussion in \cite{Rapp:1997zu} of the full  topology-induced 't Hooft Lagrangian, Fierz-rearranged to the diquark channels, but just note that for two flavors its main attractive term is the $square$ of the scalar diquark
operator
\be S_a=
\epsilon_{abc}(u^T_b C \gamma_5 d_c) \ee
where $C$ is charge conjugation and $T$ means the transposed spinor. The coefficient have the $N_c$
dependence in a simple form of $1/(N_c-1)$: which is 1 for $N_c=2$, as required by the symmetry argument
given above, and 1/2 for $N_c=3$. This simple factors confirmed finding made in the instanton liquid
framework previously \cite{Schafer:1993ra}, which concluded that the $binding$ of
the  scalar diquark is very large, comparable to a single effective quark mass $\sim 350- 400\, MeV$. 

In the last two decades there were multiple attempt to calculate properties of various diquarks 
on the lattice. Again, we do not intend to make a review of this literature, but mention one representative work
\cite{DeGrand:2007vu}, and in particular reproduce their Fig.8  here. It summarizes the dynamical quark results.
Two obvious comments on these plots are: (i) the only channel in which significant interaction (attraction)
is detected is the scalar diquark we discuss; (ii) the effect becomes more pronounced as the quark mass
diminishes. Unfortunately, in this decade old paper the quark mass is still not what it should be, but one can at least
extrapolate.
  
\begin{figure}[htbp]
\begin{center}
\includegraphics[width=8cm]{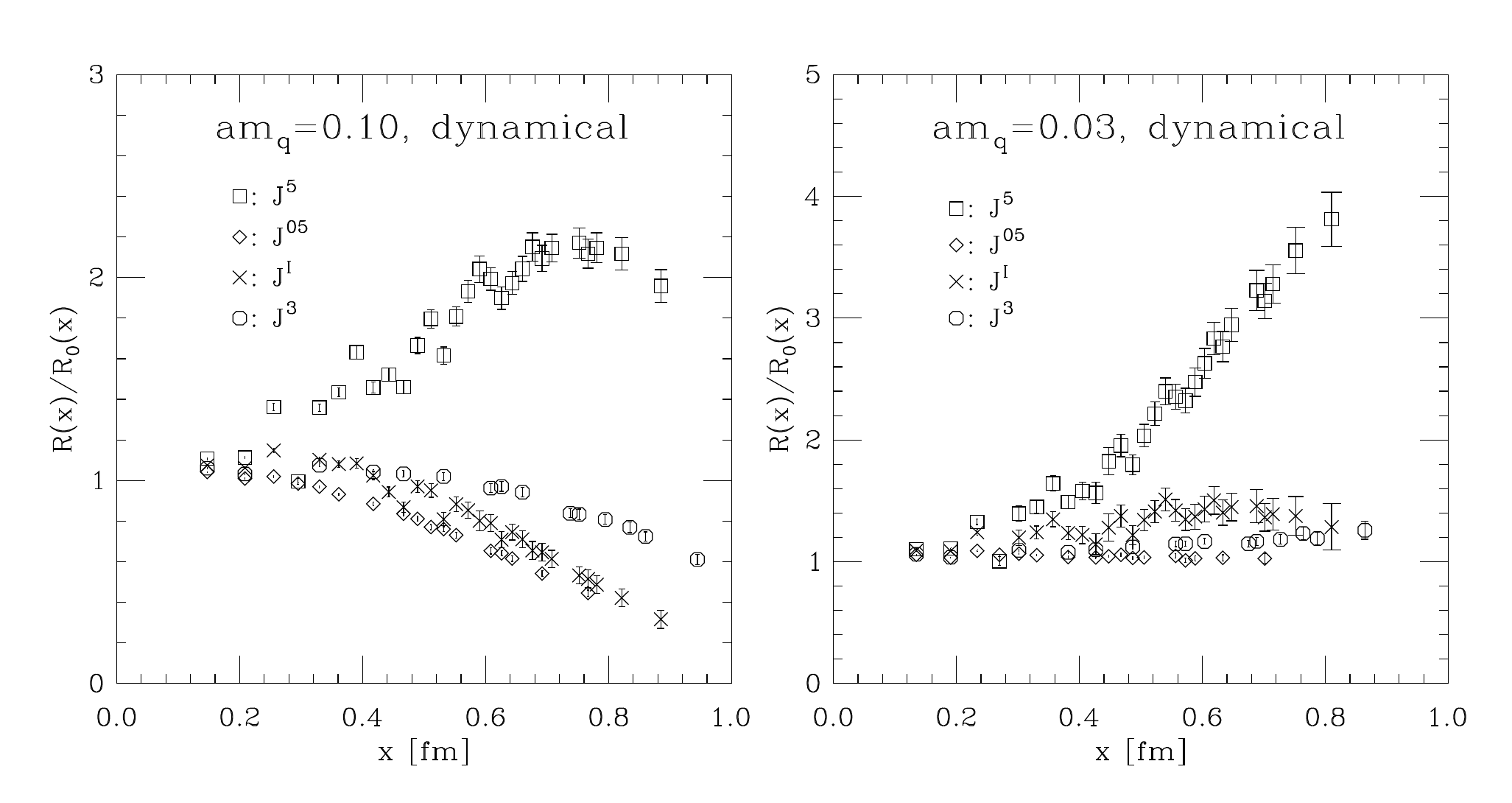}
\caption{The measured point-to-point diquark correlators, normalized to such for free massless quarks versus distance.
The open squares are for scalar diquark, others for three other diquark types. Two plots are for
different quark masses indicated in the plot.  }
\label{fig_deGrand2}
\end{center}
\end{figure}

Many other experimental measurements have indications to quark-diquark structure of the nucleon.
For example, JLAB experiment \cite{Asaturyan:2011mq} have studied intrinsic parton momenta
and found them to be very different for $u$ and $d$ quarks, qualitatively consistent with it.

However, one also needs to add a disclaimer. Applications of the diquark model
to multi-quark hadrons, such as pentaquarks $\bar s (ud)^2$
and dibaryons $(ud)(us)(ds)$ \cite{Jaffe:2003ci,Shuryak:2003zi} were based on a simple additive model .
Unfortunately, later studies \cite{Pertot}
has revealed strong repulsion between two or more diquarks containing the same quark, pushing
masses of exotic hadrons much higher than estimated in these works. Indeed, topology in vacuum is relatively dilute, and  a single
instanton can only provide  $one$ zero mode per flavor, and Pauli principle prevents binding of more than one diquark.
Since such effect should be absent for perturbative forces, the issue of multi-quark hadrons needs to be
investigated on the lattice more quantitiatively.

Now, how one can test the quark-diquark idea on the lattice?
One test  would be to study the spin structure of hadrons, with a nonzero spin but without scalar diquarks.
Such hadrons can be $\rho$ mesons or baryon decuplet particles such as $\Delta$. Unfortunately,
both are resonances with about 100\, MeV width, which complicates this approach. 

Much more feasible -- and in a way already done multiple times but not properly emphasized -- is
to systematically comparison of the couplings of the nucleon to all possible currents of the form $ \epsilon_{abc}(u^T_b C \Gamma^A  
d_c) \Gamma^A u_a$, with various spin-color matrices $\Gamma^A$.

Furthermore, one can do standard three-point correlators, measuring the expectations values of $all$
4-quark operators of the  types $\hat O_\Gamma=[\epsilon_{abc}(u^T_b C \Gamma^A  
d_c) ]^2 $ over the nucleon.  The quark-diquark model would predict that the density of scalar diquark, for
$\Gamma=\gamma_5$, would be significantly enhanced compared to  other operators. Such measurements are similar to what
was already done for weak decays of the hyperons, although with different operators.

Finally, let us repeat here some arguments of why the topology-induced effects have much better chances to explain 
the spin and isospin sea polarization than the perturbative $g\rightarrow \bar q q$ vertex. The latter
is flavor blind and of vector structure $\bar q_L q_L+\bar q_R q_R$, while the former is 
flavor asymmetric and violates chiral symmetry. As noticed in \cite{Dorokhov:1993fc}, the 't Hooft Lagnagian
would induce splitting functions for a $u_L$ quarks only of the type  \be u_L\rightarrow u_R (\bar d_R d_L) \ee
which one needs to add to the perturbative Altarelli-Parisi evolution equation. 


Let me end with a more general point. The problems of the nucleon structure and its spin
 is just a small part of the problem of the vacuum structure.  In particular,
 the diquark correlations cannot be treated in nonrelativistic or other simplified pQCD-like models 
 which do not provide  proper account for chiral $SU(N_f)$ and $U(1)_a$ symmetry breaking and pions.  
These correlations are known to  increase strongly,  as the quark mass decreases to its physical values, even for rather small masses
as seen in the plot above. The extent to which
those correlations are or are not topology related remains to be studied on the lattice in more detail.

\end{document}